# Unsupervised Feature Learning for Online Voltage Stability Evaluation and Monitoring Based on Variational Autoencoder

Haosen Yang, Robert C. Qiu, Xin Shi, Xing He

*Department of Electrical Engineering, Research Center for Big Data Engineering Technology, State Energy Smart Grid Resarch and Development Center, Shanghai Jiaotong University, Shanghai 200240, China*

**Abstract**

With the increase of uncertain elements in power systems and extensive deployment of online monitoring devices, it is necessary to search a more real-time and robust voltage stability assessment method. This study, using PMU monitoring data, explores a novel data-driven approach for long-term voltage stability assessment based on variational autoencoder (VAE). Our method is capable of extracting the most representative features by an unsupervised data mining method in a probabilistic learning way. Different from most of familiar feature extraction methods, it regularizes latent features in an expected stochastic distribution. Furthermore, a statistical indicator by sampling latent features after variance reduction is proposed to assess long-term voltage stability. Our approach is tested in various simulated power systems with different load increment models. Other cases show the accuracy and speed of our approach for estimating voltage collapse point. These testing cases successfully demonstrate the accuracy and effectiveness of our approach.

## 1. Introduction

Power system voltage stability has always been an important scientific topic, since lots of accidents caused by voltage collapse, examples including France 1978, Belgium 1982 and Japan 1987, all brought about huge economic losses. In recent years, higher penetration of sustainable energy and increasing uncertainties of system parameters have become new challenges for traditional model-based approaches to be accurately applied [1].

In addition, large amounts of operating data have been collected from broad deployment of online monitoring systems recently, especially the phasor measurement unit (PMU) with high reporting rate and accuracy. As a kind of strategic resource, PMU monitoring data encourages the development of applicable data processing

✩This work is supported by the National Key R & D Program of China under grant 2018YF-F0214705

technologies in power systems, and provides an opportunity for obtaining more real-time and precise assessment methods of voltage stability.

The classic Newton-Raphson method to solve power flow equation cannot converge when the operating point is close to voltage collapse point (VCP). So the continuation power flow (CPFLOW), using a tangent vector to predict and correct the next operating point, became the most frequently-used method to obtain a P-V curve. However, this method fails to meet the requirement of computational efficiency, thus is unable to implement in real-time fashion.

In order to realize the online monitoring of voltage stability, many researchers attempted to investigate new approaches in either model-based or data-driven way. L-index proposed by Kessel in [2] and P-index proposed in [3] are effective evaluation indicators, for their computational convenience and rigorous theoretical basics. The minimum singular value of power flow jacobian matrix [4], or the maximum singular value of inverse jacobian matrix can also represent the voltage stability level, since VCP suggests to singularity of jacobian matrix [5]. Thévenin equivalent methods, which have developed rapidly of the decade, simplifies the entire power grid from a point view of a single node into a thévenin equivalent circuit [6, 7, 8] and estimates parameters by measured data. Besides, in [9], a long-term voltage stability margin estimation method based on artificial neural network (ANN) was proposed, whose effectiveness and accuracy were demonstrated. But too much time was consumed in training and overfitting can easily happen. Literature [10] investigated the use of multi-linear regression models (MLRM) for online voltage stability assessment. In [11], the ordered singular values of PMU monitoring data matrix were mapped to the singular values of jacobian matrix through the well-known least square to indicate voltage stability. In [12], an active learning method to enhance the data preparation and training process was studied. Other approaches based on convolutional neural network (CNN) [13], random vector functional link [14] and random forest (RF) were also studied for voltage stability.

By reviewing, there are three main drawbacks: (1) The features or indicators extracted from PMU monitoring data in most of literatures can only reflect the voltage level, but are difficult to reveal the load level. For instance, the minimum singular value of monitoring data changes very slightly when load demand is distant to VCP even though the load grows greatly. Contrarily, it changes a lot when load demand is closed to VCP even if only little increment of load occurs. Therefore it is difficult to effectively characterize the changes of the load level. (2) Partial literatures used simple statistical features or manual features of monitoring data, such as mean, variance, skewness, kurtosis, eigenvalues and singular values, etc. They are sensitive to the choices of designers and not sufficiently powerful to capture the most representative features of system status. (3) Many voltage stability indicators were extracted from a prior parametric model of monitoring data, which are usually based on certain assumptions and simplifications.

Inspired by the above limitations, this paper proposes a data-driven approach for long-term voltage stability assessment based on variational autoencoder. Our method processes voltage data collected directly from PMU, and operates without any information of physical models. As a specific usage of neural network, aotuencoder is an unsupervised learning framework which extracts features naturally and automatically without any manual labels [15]. Autoencoder has already been successfully applied in the field of abnormal detection, image classification and data denoising [16].

Variational autoencoder (VAE) is a kind of enhanced autoencoder based on variational inference, aiming to extract probabilistic latent features in an expected probability distribution with parameters to be trained. It can be seen as a numerical solution to variational Bayesian problem in the condition that the correlation between latent variables and model parameters is too complex to be analytically expressed. Latent features drawn from VAE are difficult to be observed directly but show significant impact for system operation [17]. VAE has been applied in many industrial fields, including abnormal detection [18], computer vision [19] and robotics [20]. A large number of experiments have demonstrated that the representation ability of VAE is more powerful than previous methods and autoencoders [21][22]. For a P-V curve, our VAE based method is capable of drawing an approximation of the latent loading factor justly by nodal voltage data.

Main advantages of our approach are listed as follows: (1) As a data-driven approach, it conducts analysis requiring no prior knowledge of system topology, parameters and control strategies, etc. This avoids complex system modelling and calculation [23]. (2) Our method is an unsupervised learning method which does not require any labels and records. (3) Partial nodal voltage magnitudes from PMU are sufficient for our method, which accords with the fact that PMU is not installed on all nodes.

The remainder of this paper is organized as follows. Section 2 introduces the problem statement of voltage stability and details of variational autoencoder. Section 3, with the aid of a tutorial example, introduces our VAE based method for voltage stability assessment. Section 4 includes numerous cases, and section 5 is conclusion.

## 2. Problem Statement and Variational Autoencoder

### 2.1. Voltage Stability in Power Systems

The P-V curve, plotted by the variation of nodal voltage magnitudes with respect to power demand and generation outputs, is the most frequently-used methodology to assess voltage stability [24]. As shown in Fig. 1, by tracing a P-V curve, the relationship between load and voltage magnitudes is clearly expressed. Each point in a P-V curve accords with an operation state that amenable to power flow equation, which can be expressed as below:

$$P_i = p_i(\theta_1, ..., \theta_N, V_1, ..., V_N), i \in N_{PQ} \cup N_{PV}$$
$$Q_i = q_i(\theta_1, ..., \theta_N, V_1, ..., V_N), i \in N_{PQ} \tag{1}$$

where $P_i$, $Q_i$, $V_i$, $\theta_i$ denote the active power, reactive power, voltage magnitude and voltage angle of node $i$, respectively. $N_{PQ}$ and $N_{PV}$ denote the set of PQ nodes and set of PV nodes. $p_i(\ )$ and $q_i(\ )$ are nonlinear functions. Besides the power flow equation, the profile of a P-V curve is also dominated by load changing models. For a certain grid, different load increment models or directions generate various P-V curves, and subsequently lead to different stability criterions. Notice that load increment model is an assumption, depending on operators or historical data. In general, load demand and generator outputs in different nodes increase at different rates. So the increment manner can be modeled as:

$$\begin{aligned} P_i &= P_{i0}(1 + \lambda k_{P\,i}) \\ Q_i &= Q_{i0}(1 + \lambda k_{Qi}) \\ G_j &= G_{j0}(1 + \lambda k_{Gj}) \end{aligned} \quad (2)$$

where $P_i$, $Q_i$ denote the active power and reactive power demand for load $i$, $P_{i0}$, $Q_{i0}$ suggest the load demand of current status. $G_j$ represents the output for generator $j$, while $G_{j0}$ is the generator output in basic case. $k_{P\,i}$, $k_{Qi}$, $k_{Gj}$ are multiplicative factors denoting different increasing rates. $\lambda$ is the critical loading parameter defining the variation of load demand and generator outputs on a certain direction.

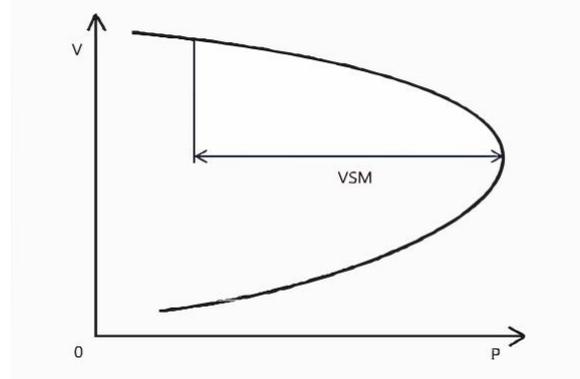

Figure 1: P-V curve and voltage stability margin.

When load increases, voltage magnitudes will drop down at the same time the operating point will move close to VCP. The VCP works as a boundary separating stability and instability domains, and power grid accidents will occur when it operates below VCP [25]. VCP is defined as the point at the maximum of loading factor:

$$VCP = (\lambda_{max}, V_i^{\lambda_{max}}) \quad (3)$$

where $\lambda_{max}$ is the maximum of loading factor $\lambda$, $V_i^{\lambda_{max}}$ is the corresponding voltage magnitude of node under concern. Power grids are supposed to operate far away from VCP to ensure adequate margin of stable operation

### 2.2. Data of Voltage Stability

Nodal voltages, as the solution of power flow, are the most important variables in power systems. With the gradual deployment of PMU, the phases and magnitudes of nodal voltages can be measured directly and precisely. But there is also a problem that it is difficult to install PMU on every node due to the high expense. However, our method based on VAE, is tolerant for this incomplete data collection. In this paper, voltage magnitudes and angles $V_{in} = [V_1, V_2...V_n, \theta_1, \theta_2...\theta_n]$ sampled by PMU are analyzed. A split window is used to sample a period of PMU monitoring data. For a series of time $T$, by arranging these vectors $V_{in}$ in chronological order, a spatio-temporal data matrix can be obtained $X = [V_{in}^1, V_{in}^2, ..., V_{in}^t]$, where $V_{in}^t$ is the transpose of the current PMU monitoring vector.

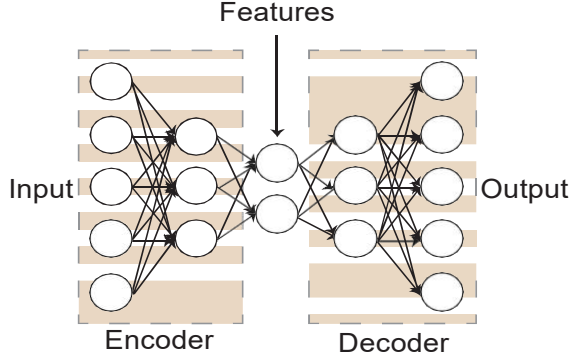

Figure 2: Autoencoder model.

The reasons why we select voltage measurements of PMU are listed as follows: (1) Although power flow and injected power are measurable by SCADA, the accuracy and measuring frequency are a lot worse than that of PMU [26]. So the voltage data collected by PMU ensures the high accuracy of voltage stability analysis. (2) Nodal voltages are the most basic variables which are sensitive to the change of operating status in a power system. (3) Nodal voltages are the solution of power flow. If power flow calculation is regarded as a black box, nodal voltages are the outputs which are supposed to represent the overall system status.

*2.3. Autoencoder*

Autoencoder is an unsupervised learning approach which extracts nonlinear features automatically. As an auto-associative neural network, it is designed to reconstruct the input space in the output space after a projection in a low-dimensional space [27]. Therefore, what should be concerned is the hidden layer instead of the final outputs. The outputs of middle hidden layer of an autoencoder are regarded as the feature hypothesized to show important properties of the whole system [28]. The multi-layer neural networks before $z$ are known as encoder, while networks after $z$ are named decoder, as shown in Fig. 2. Calculations of every layer in networks can be written as:

$$y = f(Wx + b) \qquad (4)$$

where $W$, $b$, $x$, $y$ respectively denote the weights matrix, bias vector, input vector and output vector of a layer. $f(\ )$ denotes the activation function.

The encoder is responsible for extracting features while the decoder is in charge of reconstructing the origin data from hidden feature $z$ [29]. The reconstruction mean square loss (MSE) is:

$$Loss = \|x - \tilde{x}\|_2 \qquad (5)$$

where $\tilde{x}$ denotes the reconstruction data. As same as simple neural networks, training of autoencoders still depends on the gradient descent method to update weights and bias in every layer iteratively. The latent feature $z$ is low-dimensional and representative for the entire data set, provided that the reconstruction loss has already been reduced properly.

## 2.4. Variational Autoencoder

The above mentioned autoencoder has various possible latent features satisfying the condition of small reconstruction loss. In other words, low-dimensional feature $z$ varies randomly with different initializations and hyper parameters, i.e., feature $z$ will be different if the model is retrained. Therefore, in this paper as a probabilistic learning method, VAE is utilized to learn a probabilistic feature. VAE is an enhanced autoencoder that formulates variational Bayesian problem as a neural network. In VAE, features are designed to follow an expected probability distribution with parameters to be tuned, and the encoder is exactly used to output parameters of the distribution, as shown in Fig. 3. The hidden feature $z$ is a sample from the expected distribution, while the decoder is responsible for reconstructing origin data from $z$. Hence VAE requires an approach to compare the stochastic distribution of $z$ with the expected distribution. KL divergence is widely used to measure the difference between two stochastic distributions, which is:

$$KL(p//q) = E_{p(x)} \left[ log \frac{p(x)}{q(x)} \right] \tag{6}$$

where $p(x)$ and $q(x)$ are two compared probability density functions.

In Bayesian method, the conditional distribution of the encoder and the decoder can be written as a posterior $p(z|x)$ and a likelihood $p(x|z)$ respectively, while we denote the expected posterior by $h(z|x)$ (Gaussian, Bernoulli distributions, etc.). Considering $p(z|x) = p(x, z)/p(x)$, the KL divergence between $p(z|x)$ and $h(z|x)$ can be rewritten as [30]:

$$KL(p(z|x)//h(z|x)) = E_{p(z|x)}[log p(z|x)] \\ - E_{p(z|x)}[log h(x, z)] + log h(x) \tag{7}$$

where $log h(x)$ is the log-likelihood of reconstructed data, $log h(x, z)$ is the joint probability density, and $log h(x) = E_{p(z|x)}[log h(x, z)]$. VAE uses an elbo (evidence of lower bound) loss function whose definition is:

$$elbo = E_{p(z|x)}[log h(x, z)] - E_{p(z|x)}[log p(z|x)] \tag{8}$$

The divergence is always greater than or equal to zero, implying that maximizing the elbo shares the same effect as minimizing the KL divergence. By unfolding the first term of (8), we can obtain:

$$elbo = E_{p(z|x)}[log h(x|z)] - KL(p(z|x)//h(z|x)) \tag{9}$$

It is intractable to calculate the first term of (9) directly. However, $h(x|z)$ can be treated as the process of yielding the distribution of reconstructed data from latent feature $z$, while the expected distribution $h(z|x)$ is regarded as an approximated posterior distribution of $z$ [31]. The first term $E_{p(z|x)}[log h(x|z)]$ can be estimated by

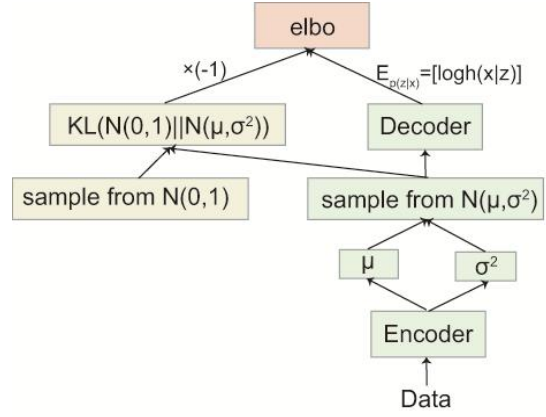

Figure 3: VAE process.

cross entropy between the output of the decoder and data:

$$E_{p(z|x)}[\log h(x|z)] \because -E_{p(x|z)}\log p(x) \\ = -\sum_{i=1}^{n} \hat{x}_i \log x_i + (1-\hat{x}_i)\log(1-x_i) \quad (10)$$

where $\hat{x}_i$ is the *ith* output of the decoder. Generally, standard Gaussian distribution is reasonable to model $\log h(x|z)$ [30]. A multivariate Gaussian distribution has two groups of parameters: mean vector $\mu(z)$ and diagonal variance matrix $\sigma^2(z)$. Thus the encoder is designed to have two parts of outputs from one network, as shown in Fig. 3.

As for the second term in (10), KL divergence between two Gaussian distributions can be written as:

$$KL(\mathcal{N}(\mu, \sigma^2) // \mathcal{N}(0, 1)) = \frac{1}{2}(-\log\sigma^2 + \sigma^2 + \mu^2 - 1) \quad (11)$$

What's more, the training of VAE needs an advanced optimization. Adam algorithm, a first-order stochastic gradient descent method based on adaptive estimation of lower-order moments, is well known for its effectiveness recently [32].

## 3. Unsupervised Learning for Voltage Stability

Now we show how the feature $z$ is used to assess voltage stability. To facilitate reading, a brief tutorial example is conducted to help us clarify the methodology.

*3.1. Feature Learning in Voltage Stability Assessment*

As shown in function (1), a vector containing voltage magnitudes and angles $V_{in} = [V_1, V_2...V_n, \theta_1, \theta_2...\theta_n]$ is the solution of power flow with inputting nodal injective active and reactive power $L_{in} = [P_1, P_2...P_n, Q_1, Q_2...Q_m]$. Since the load vector $L_{in}$ is not ordinarily accessible to PMU, for real-time voltage stability assessment, $L_{in}$ can be regarded as unobserved latent variables inside the voltage vector $V_{in}$.

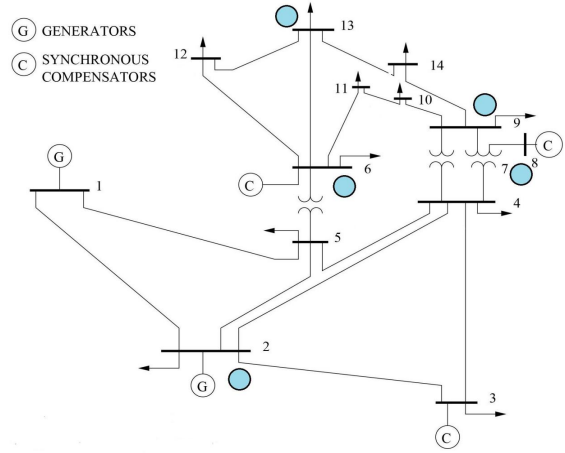

Figure 4: IEEE 14-bus system, the nodes with PMU installation are marked by blue circles.

But directly making an approximation for all entries in $L_{in}$ must lie on the complicated physical model that makes the algorithm very time-consuming, and thus not real-time sufficiently. However, for voltage stability, there is no requirement to approximate all load demand, but only to pursue some crucial features that govern the load changing trends. So the task of our study is searching low-dimensional features from voltage measurement, which can also properly indicate the latent load profile.

This problem is a probabilistic approximation issue under the scenario with latent variables. Traditional methods for approximating latent variables mainly include expectation-maximization (EM), Markov chain Monte Carlo (MCMC) and Variational Bayesian algorithm [33]. These methods, however, unanimously require explicit relationship between latent variables and measurement data for iterative optimization, which is computationally intractable for voltage stability since the nonlinear power flow equation of power systems is extremely complex. Therefore, the VAE, a numerical way analogous to variational Bayesian method, is utilized to extract crucial yet low-dimensional features of voltage stability. VAE shares the same optimization objective with variational Bayesian approach, while an encoder and a decoder are used to substitute the complicated physical model. The values extracted through VAE are not only representative features of inputting data like other nonlinear dimension reduction methods, but also latent variables which can regenerate the probability distribution of the dataset. Beneath it all, we use a black-box inference method, the VAE, to extract dominate features in low-dimensional space in a probabilistic way.

For simplicity, the tutorial example assistant for explanation employs IEEE 14-bus system, and the nodes with PMU installation are marked by blue circles in Fig. 4. The detailed structure and hyper parameters of our VAE network are listed as follows:

(1) Length of features: 2. So there are four parameters (mean and variance of every feature).

(2) Layers of encoder and decoder: 3.

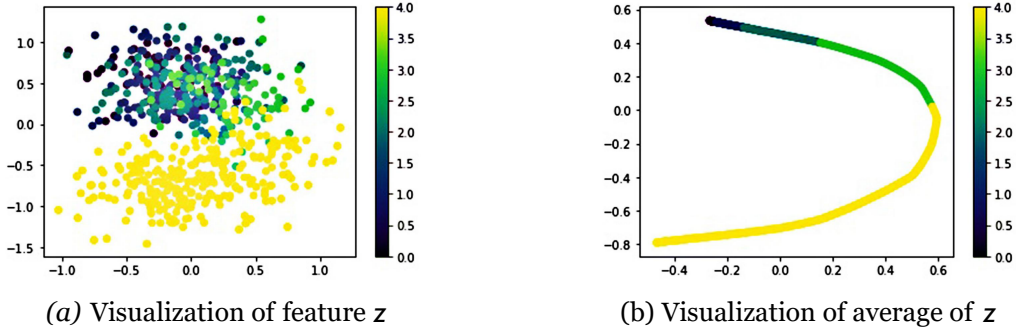

*(a)* Visualization of feature *z*     (b) Visualization of average of *z*

Figure 5: visualization of features. The abscissa and ordinate are the two features, which is [($z_1$), ($z_2$)] in (a), and [$E(z_1)$, $E(z_2)$] in (b). This tutorial example is tested in IEEE 14-bus system. The active power of node 4 increases monotonically. A time series of voltage measurement on a P-V curve simulated by CPFLOW are input successively, every measured point on this P-V curve yields a point in this figure. Black, dark blue and green points denote 87.5%-100%, 75%-87.5%, 50%-75% distance from the initial operating point to VCP, respectively.

(3) Number of neural units of encoder: 100, 100, 4 (four parameters of the expected probability distribution).
(4) Number of neural units of decoder: 100, 100, 10 (size of the output vector).
(5) Activation function: Relu. Except the activation function of variance output unit in middle layer is softplus.
(6) Gradient descent method: Adam.
(7) Learning rate: 0.0001.

The initialization of network parameters is stochastic samplings from standard Gaussian distribution $N(0, 1)$. After adequate offline training, VAE model is applied to online monitor voltage stability. To thoroughly yet simply show the characteristics of our method, we choose a simplex load increment model that one active load increases monotonically, but draw the full P-V curve in the tutorial case.

The latent feature *z* of every point on an entire P-V curve is shown in Fig. 5(a), in which different colour denotes different risk level. Black, dark blue and green points denote 87.5%-100%, 75%-87.5%, 50%-75% distance from the initial operating point to VCP, respectively, while yellow points denote operating points below VCP. The feature *z* is a 2-value vector according to the experience. It is clear that features shown in Fig. 5(a) are a stochastic sequence instead of deterministic values because they are samplings from $N(\mu(z), \sigma^2(z))$. Then we spontaneously consider to sample them after variance reduction. We introduce a temperature parameter $\phi$ to modify the level of randomness:

$$\hat{\sigma}_i^2 = \sigma_i^2 \phi \qquad (12)$$

where $\phi$ is set between 0 and 1, and $\sigma_i^2$ is the variance of the *ith* feature $z_i$. We use $\hat{z}$ to denote the samples after variance reduction, i.e., $\hat{z} \sim N(\mu(z), \hat{\sigma}^2(z))$. The $\hat{\sigma}_i^2$ will become the average of $z_i$ if $\phi$ is equal to zero. Fig. 5(b) shows the average of our feature [$E(z_1), E(z_2)$].

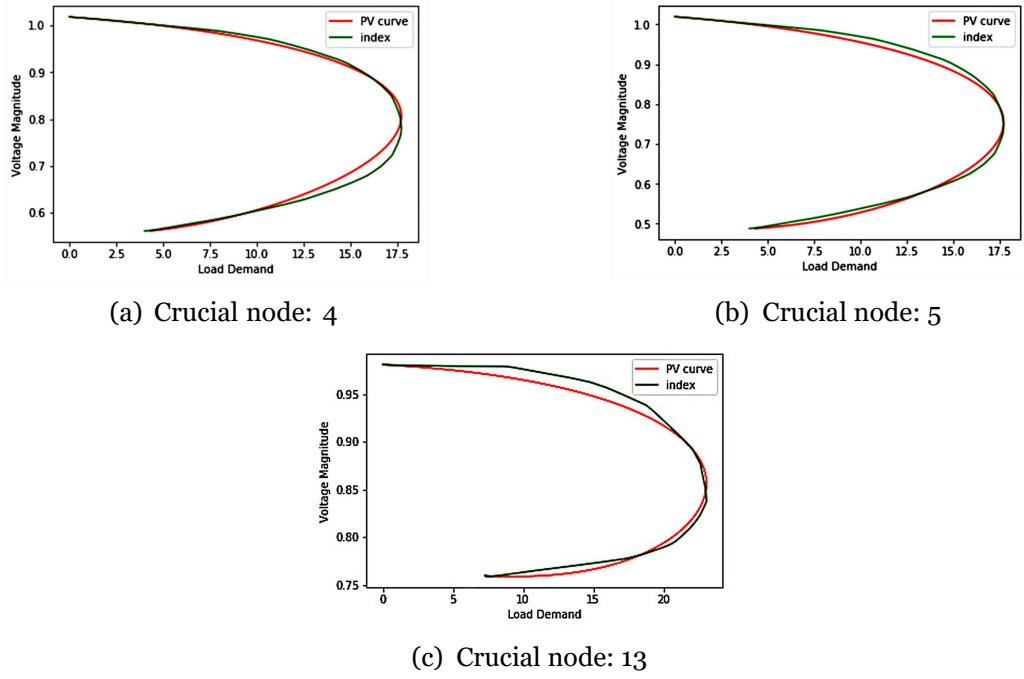

(a) Crucial node: 4

(b) Crucial node: 5

(c) Crucial node: 13

Figure 6: Comparison between our index and P-V curve. Green line is the curve of our index, red line is the real P-V curve. Every subfigure chooses a single crucial node whose active power increases until the whole P-V curve is traced. To make them comparable, the features are linear transformed by multiplying coefficient vector $\beta$.

One of the most effective aspect of our approach is that the variation of our indicator $[\hat{z}_1, \hat{z}_2]$ has nearly the same shape as a P-V curve. The first extracted feature $\hat{z}_1$ can approximate the loading parameter $\lambda$, while feature $\hat{z}_2$ reflects to voltage magnitudes. In other words, by analyzing the solution (Voltage) of CPFLOW, we can approximate a representation of the input (load demand), which agrees with our analysis above. This can be explained by that our VAE aims to find the most crucial latent variables of data. For voltage stability problem, the loading parameter $\lambda$ is the most crucial variable, which dominates the change direction of the entire system. And the overall degree of voltage is another crucial feature which represents the operating status. What should be noticed is that the second feature represents the overall voltage amplitude of the whole system, rather than the voltage magnitude of a certain node, since a simple load increase tends to cause decrease of multiple nodal voltage magnitudes. For example, in this case, the increase of active power of node 4 causes decrease of voltage of node 4, 5, 7, 9, 10, 11, 12, 13, 14. Even though only voltage data is analyzed, our method is capable of approximating the loading increment trends, because of the powerful ability of probabilistic feature learning.

*3.2. Comparison with Real P-V curve*

In this subsection, we compare our obtained curve with the real P-V curve. To achieve the comparison, a least square method is used to transform our feature $[\hat{z}_1, \hat{z}_2]$ to the real P-V curve. For a feature point $\hat{z} = [\hat{z}_1, \hat{z}_2]$, we denote a time series of features by $Z_i = [\hat{z}_{(1)}, \hat{z}_{(2)}....\hat{z}_{(T)}] \in R^{T*2}$. And a series of loading factors and voltage magnitudes are denoted by a matrix $C_i = [\lambda_{(1)}, \lambda_{(2)}...\lambda_{(T)}; v_{(1)}, v_{(2)}, ..., v_{(T)}] \in R^{T \times 2}$. Then multiple time series of features and P-V curves are gathered to form a matrix respectively: $Z = [Z_1, Z_2, ..., Z_n] \in R^{N T \times 2}$ and $C = [C_1, C_2, ..., C_n] \in R^{N T \times 2}$, where $N$ is the number of time series. Then the least square problem is:

$$\beta = argmin ||\beta^T Z - C||_2 \qquad (13)$$

where $\beta$ is the coefficient. The solution of this problem is:

$$\beta = (Z^T Z)^{-1} Z^T C \qquad (14)$$

The comparison between $\beta^T Z_n$ and $C_n$ is shown in Fig. 6. Each subfigure is drawn with a crucial load increasing. The crucial node of Fig. 6(a), Fig. 6(b) and Fig. 6(c) are node 4, 5 and 13, respectively. The comparison results suggest that our feature approximates P-V curve well. Therefore, our feature automatically extracted from PMU data is effective to assess voltage stability. And the greatest advantage is that our feature is similar to P-V curve since P-V curve is the most straightforward expression of the relationship between power and voltages, by which we can acquire the load level clearly of an operating state. And certainly, there is inevitable error between features and the real P-V curve due to the random samplings, measuring noise and incomplete information.

---
**Algorithm 1** Offline Training Process of Proposed Approach
---
1: Build the VAE network as shown in Fig. 3 and determine hyper parameters by experience;
2: Randomly assign the weight matrix $W$ and bias $b$ by sampling from standard Gaussian distribution $\mathcal{N}(0, 1)$;
3: **While** not converge **do**

   1. Randomly select a group of segments of P-V curves from training dataset;
   2. Use elbo-oriented loss function (9) and Adam algorithm to update all the weights and bias;

4: **End While**

---
**Algorithm 2** Online Operation of Proposed Approach
---
1: Collect truncated real-time voltage measurements $V_{in}$ from PMU;
2: Input $V_{in}$ into VAE models which have been fine-trained offline;
3: Collect the probability distribution of latent feature $z$;
4: Reduce the variance of feature $z$ by function (12);
5: Sample feature $\hat{z}$ from the distribution after variance reduction;
6: Apply pre-trained least square model to linear transform feature $\hat{z}$ to $\beta^T \hat{z}$, and use it to monitor voltage stability;

---

*3.3. Training*

Before the real-time operation, our VAE network needs be properly trained offline. That is, we utilize historical or simulative measurement data to update the parameters involved in VAE offline, afterwards the fine-trained VAE network can be applied online. The procedures of offline training and online operation are outlined in Algorithm. 1 and Algorithm. 2 respectively. Notice that the offline training spends a lot of time but the online operation is computationally efficient.

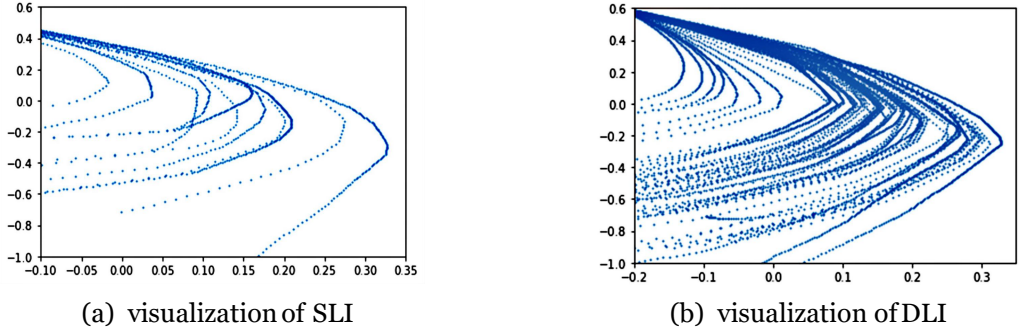

(a) visualization of SLI  (b) visualization of DLI

Figure 7: Visualization of our estimation with different load increment models of the tutorial example. (a) is the results in the condition of single load increment model that every line denotes the increment of active power of one PQ node. (b) is the results of double load increment model and every line denotes the increment of active power of arbitrary two PQ nodes in IEEE 14-bus system. The x-axis and y-axis are $\hat{z}_1$ and $\hat{z}_2$.

In this tutorial example, the training set of our VAE based method is randomly generated. We randomly select 1 to 3 nodes, of which active power or reactive power simultaneously increases until a whole P-V curve is obtained. By repeated simulation, we obtain 40 complete P-V curves to compose the training set. Since our VAE network is dedicated to understand the insightful operating mechanism of a power system instead of the load increment model, after training, our network is capable of extracting the most representative features no matter how load grows.

To compare the performance between different load growth models, we visualize the features of all single load increment (SLI) models and all double load increment (DLI) models in IEEE 14-bus system. As shown in Fig. 7(a), every curve of features corresponds to the case a single injective active power increases alone. While in Fig. 7(b), every curve of features is that arbitrary double injective active power increases simultaneously. For comparison, every feature point in these curves is drawn by the same changing step of the loading factor.

With the system changing, the procedure of our method is consistent but we need modify the hyper parameters determined manually before training. Main rules of adjusting hyper parameters are roughly listed as follows: (1) With the increase of system scale (namely, the input voltage vector enlarging), VAE needs more neural layers and more neural units in each layer. (2) With the increase of system scale and complexity, the learning rate of offline training should be turned up, since larger scale and complexity make the parameters updating of VAE become slower and more difficult. Thus larger learning rate is more prone to converge efficiently. (3) With the increase of system scale, the offline data pool for training is required to be enlarged, because a larger system needs more data to make the proposed method thoroughly understand the voltage stability mechanism of it.

*3.4. More Discussions and Comparison*

The voltage data collected from PMU must contain sufficient information of the load level and changing trends, and certainly contains multiple noise. By modeling latent features in a probability function, our method can effectively reduce noise interference and find more generalized features, because the optimization w.r.t. probabilistic features shares the similar effects with adding regularization in loss function [34]. Besides, the final outputs are also modeled in a Gaussian or Bernoulli distribution, that is, VAE pays more attention in general representation of the entire data set rather than purely pursuing the smallest loss function. Therefore, compared with other deterministic feature learning methods including autoencoder, sparse autoencoder (SAE) and deep belief network (DBN), features of our method are more representative. As shown in Fig. 8, the features extracted from autoencoder, SAE and DBF can not clearly show the characteristics of different operation states.

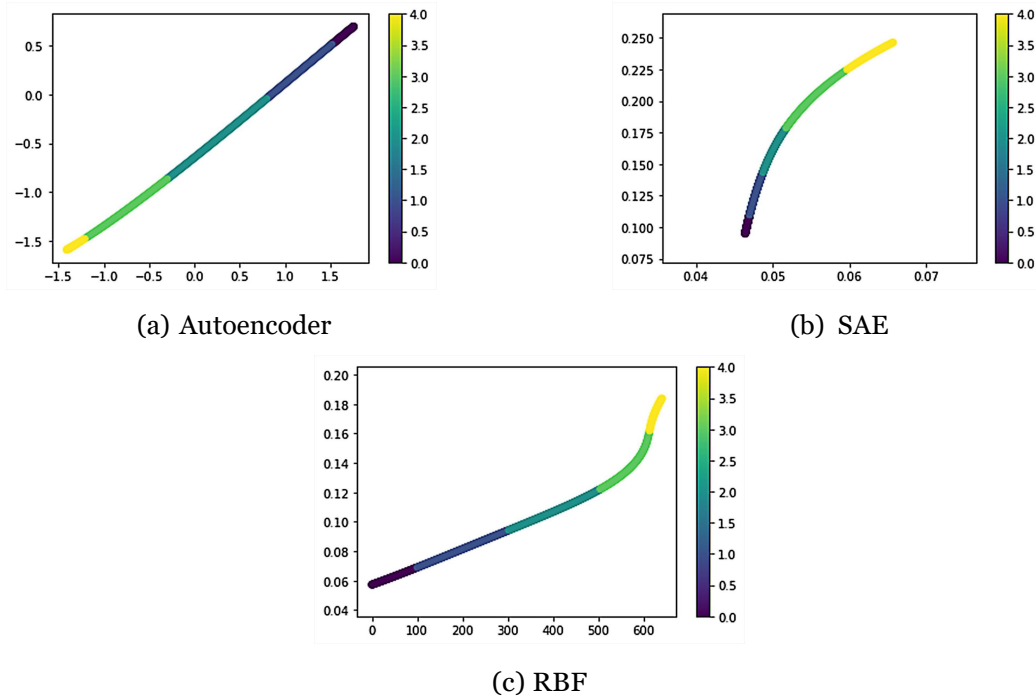

(a) Autoencoder   (b) SAE

(c) RBF

Figure 8: Feature visualization of other deterministic feature learning methods. The explicit parameters of simulations are the same as Fig. 5.

As discussed so far, the differentiation of our method from other works both in model-based and data-based way can be listed as follows. With respect to most of approaches based on learning algorithms [9, 10, 14, 35]: (1) The proposed method produces indices in an unsupervised way. This freeness of supervision circumvents the manual labelling which consumes massive computational resources and introduces uncertainties. (2) Our features for monitoring show the changing profile both of voltage magnitudes and load demand even if only real-time voltage phasor data is processed. This property is attributed to that the VAE involved is not simply feature extraction but also an approximated inference of the causal relationship inside data. (3) Our method models the probability distribution of features and inputting data instead of modelling features directly. This probabilistic modelling makes our method less affected by random measurement noise and easier to be generalized [36]. Besides, compared with model-based methods [2, 3, 4, 5, 11], the differentiation is evidently the avoidance of complex system modelling, which averts the error accumulation of multiple simplifications and estimations of parameters.

Table 1: Load changing set: Active power (MW) of node 9.

| Time | Active Power |
|---|---|
| 1-500 | 121 |
| 501-700 | 121+($t$-500) |
| 701-900 | 321 |
| 901-1200 | 321-($t$-900) |
| 1201-1500 | 121 |

## 4. Case Studies

In this section, the accuracy and effectiveness of the proposed model are validated explicitly via simulations in IEEE 57-bus, 118-bus and European 1354-bus high voltage transmission network[1]. The first three cases show the performance for real-time monitoring of voltage stability with different operating status. Besides, the accuracy of VCP estimation is also tested in the last case. The simulation of dataset is operated by MATPOWER 6.0 (runcpf.m and runpf.m), while the VAE algorithm is conducted by Tensorflow package [37] in Python. The reporting rate of PMU is presumed to be 50Hz, and other settings of PMU follow $C37.118 - 1$ standard.

*4.1. Case in IEEE 57-bus system*

This case shows the performance of our method when load demand changes in various ways. The synthetic data was sampled from the simulation of IEEE 57-bus system. In this case, gradual changes and sudden changes of load demand are simulated. The explicit load changes are shown in Tab. 1. And a little white noise was added to represent random fluctuations. The PMU placement involved is based on the weighted least square method proposed in [38]. This weighted least square method seeks a placement that deploys the minimum number of PMUs while ensuring the observability. And we then added some PMUs in other nodes with relatively more connection to maintain certain redundancy. The nodes with PMU installed are 1, 2, 6, 10, 12, 19 ,22, 24, 25, 27, 32, 36, 38, 41, 45, 46, 48, 49, 52, 55, 57. The detailed structure and hyper-parameters are listed as follows:
(1) Length of features: 2.
(2) Layers of encoder and decoder: 3.
(3) Number of neural units of encoder: 300, 300, 4.

---
[1]The explicit parameters of IEEE model refer to the embedded scripts in Matpower 6.0 (including case14.m, case57.m, case118.m, case1354pegase.m) or website: "https://icseg.iti.illinois.edu/power-cases/".

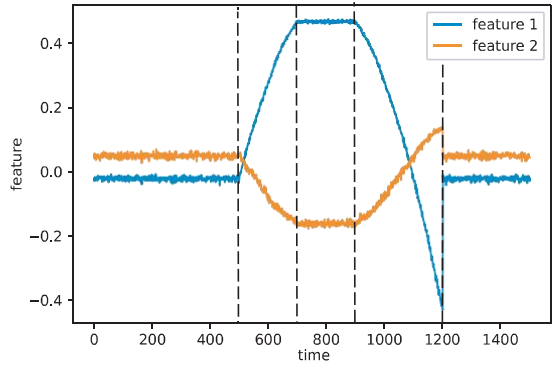

Figure 9: Features in IEEE 57-bus case.

(4) Number of neural units of decoder: 300, 300, 42.
(5) Activation function: Relu. Except the activation function of variance output unit in middle layer is softplus.
(6) Gradient descent method: Adam.
(7) Learning rate: 0.0001.
(8) Temperature parameter: 0.05.

The training set, containing voltage data of 50 full P-V curves, is also generated randomly. Each P-V curve is obtained by randomly selecting 1 to 5 nodes, of which active power or reactive power simultaneously increase, as same as the tutorial example.

As shown in Fig. 9, during $t_s = 1 \sim 500$, there is only little fluctuations in our two features since the load level remains constant and random samplings play a dominant role. During $t_s = 501 \sim 700$, the active power of node 9 linearly increases from 121 MW to 321 MW, leading to that our first feature $\hat{z}_1$ increases from $-0.0276$ to $0.4552$ and the second feature $\hat{z}_2$ decreases from $0.0526$ to $-0.1632$ gradually. Recalling our analysis in above, the first feature $\hat{z}_1$ can represent the loading factor in a similar way, and another feature $\hat{z}_2$ is capable of reflecting the overall voltage level. So the results are in line with our analysis of two features.

During $t_s = 701 \sim 900$, the active power of node 9 remains 321 MW, so the features only have minor fluctuation due to measurement noise. During $t_s = 901 \sim 1200$, contrary to $t_s = 501 \sim 700$, load demand decreases gradually from 321 MW to 21 MW, which results in decrease of $\hat{z}_1$ and increase of $\hat{z}_2$. Furthermore, $\hat{z}_1$ at $t_s = 1200$ is lower than the initial value, while $\hat{z}_2$ at $t_s = 1200$ is greater, according with the fact that the load level at $t_s = 1200$ (21 MW) is lower than the initial load (121 MW).

At $t_s = 1200$, a sudden increase of load demand is set to prove the robustness of our method for dramatic changes. Active load suddenly increases from 21 MW to 121 MW. And our features, corresponding to load demand, significantly and rapidly response, including the first feature $\hat{z}_1$ increasing from $-0.4572$ to $-0.0268$, at the same time $\hat{z}_2$ changing from $0.1130$ to $-0.1605$.

During $t_s = 1201 \sim 1500$, the load level returns to the beginning value. Our features revert back to the initial values at the same time, showing that our method

Table 2: Load changing set: Active power (MW) of node 11, 14.

| Time | Node 11 | Node 14 |
| --- | --- | --- |
| 1-500 | 70 | 10 |
| 501-800 | 70+($t$-500) | 10 |
| 801-900 | 70+($t$-500) | 10+($t$-800) |
| 901-1500 | 470 | 110 |

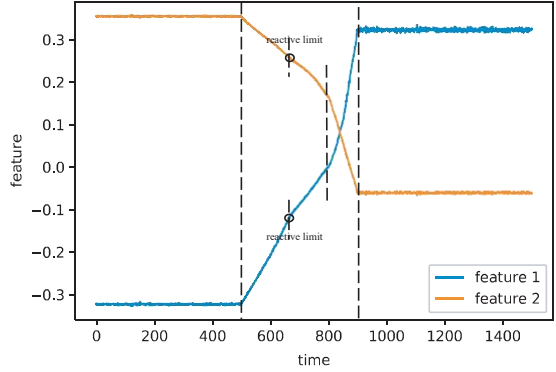

Figure 10: Features in IEEE 118-bus case.

is independent of time steps and historical data.

### 4.2. Case in IEEE 118-bus system

This case proves the effectiveness of our method when multiple load changes simultaneously in IEEE 118-bus system. The detailed settings of load change are shown in Tab. 2. Similarly, the PMU placement is based on the weighted least square with adding some PMUs to ensure certain redundancy. The nodes with PMU installed are 2, 4, 5, 9, 11, 12, 15, 17, 21, 24, 25, 28, 34, 37, 40, 45, 49, 52, 54, 56, 62, 63, 66, 68, 73, 75, 77, 80, 82, 85, 86, 89, 90, 94, 101, 105, 107, 110, 114. Hence there are 78 values in a sample vector consisting of 39 nodal voltage magnitudes and 39 nodal voltage angles. Meanwhile, the detailed parameters of our VAE network are listed as follows:

(1) Length of features: 2.
(2) Layers of encoder and decoder: 4.
(3) Number of neural units of encoder: 200, 650, 650, 4.
(4) Number of neural units of decoder: 650, 650, 200, 78.
(5) Activation function: Relu. Except the activation function of variance output unit in middle layer is softplus.
(6) Gradient descent method: Adam.
(7) Learning rate: 0.0002.
(8) Temperature parameter: 0.05.

The training set, containing voltage data of 80 entire P-V curves, is also generated randomly, as same as the tutorial example and case 1.

Table 3: Abnormal Change of nodal active power. $P_N$ is the normal active power of node $N$. During $t_s = 801$~$850$, 30 nodes are randomly selected for sudden change.

| Time | Changing | Node |
|---|---|---|
| 1-400 | No change | / |
| 401-700 | $(1+(t-400)/1200)P_N$ | All |
| 701-800 | $1.25P_N$ | All |
| 801-850 | $1.28P_N$ | 30 Nodes |
| 851-1000 | $1.28P_N$ | All |
| 1001-1500 | $(1.28-(t-1000)/1500)P_N$ | All |

As shown in Fig. 9, during $t_s = 1$ ~ $500$, there are only little fluctuations in our two features because of the same reason. During $t_s = 501$ ~ $800$, the active power of node 11 increases solely from 70 MW to 370 MW. And at $t_s = 669$, generator on node 8 reaches its reactive limit, so that the increasing or decreasing rate of two features changes, demonstrating that our features is sensitive to the reactive limit.

During $t_s = 801$ ~ $900$, active power of node 14 also increases simultaneously, causing a more rapid change of our two features. The more rapid the overall load demand changes, the quicker our features change, showing that our features are sensitive to the rate of load change.

In summary, through these cases and the tutorial example, our method is sensitive to load demand, voltage level, changing rate of load demand, reactive power limit, and is independent to time steps and historical data.

*4.3. Case in European 1354-bus High Voltage Transmission Network*

This case provides the results of our method in a larger system with a more realistic operation situation. The European high voltage transmission network is employed, which contains 1354 buses, 260 generators and 1991 branches [39]. Different from the cases exhibited above, this case involves simultaneous changes of all load in either continuous or occasional way. The detailed settings of load change are shown in Tab. 3. Similarly based on the weight least square method, 478 PMUs at total are placed. The explicit hyper parameters of our method are as follows:

(1) Length of features: 2.
(2) Layers of encoder and decoder: 5.
(3) Number of neural units of encoder: 1500, 1500, 1100, 700, 300, 2.
(4) Number of neural units of decoder: 300, 700, 1100, 1500, 1500, 956.
(5) Activation function: Relu. Except the activation function of variance output unit in middle layer is softplus.
(6) Gradient descent method: Adam.
(7) Learning rate: 0.0004.
(8) Temperature parameter: 0.05.

As shown as Fig. 11, the first change occurs at $t_s = 400$ and all active load begins to increase at the same rate (2.5% per second). This simultaneous and monotonic increment leads to that feature $\hat{z}_1$ increases from -0.2332 to 0.3289 and $\hat{z}_2$ decreases from 0.20879 to -0.082, amenable to the theoretical analysis of our approach. At $t_s$ = 800, an occasional increment of randomly selected 30 load happens. However, this sudden ascent fails to cause a large change of features, since the amplitude of load increment is not large sufficiently.

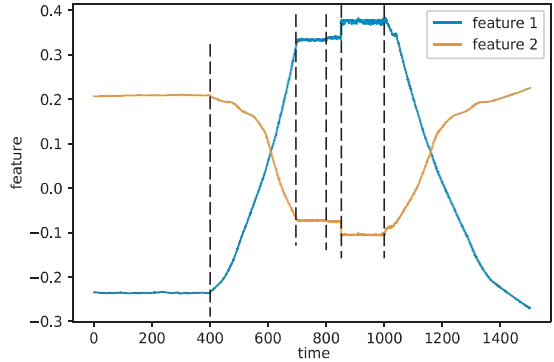

Figure 11: Features in European 1354-bus high voltage transmission network.

Table 4: Operation Speed.

| System | Input size | time(ms) |
|---|---|---|
| IEEE 14-bus | 10 | 0.0020 |
| IEEE 57-bus | 42 | 0.0062 |
| IEEE 118-bus | 78 | 0.0129 |
| European 1354-bus | 956 | 0.0265 |

At $t_s$ = 850, all load rises suddenly to the level of 1.28$P_N$, where $P_N$ is the normal level of active power at node $N$ (normal levels are referred to script case1354pegase.m in Matpower 6.0 or [39]). Accordingly, feature $\hat{z}_1$ ascends and feature $\hat{z}_2$ descends, illustrating that our approach can also adapt to large systems with simultaneous and sudden load increment. During $t_s$ = 1000~1500, all load drops simultaneously and gradually, resulting in corresponding descent or increase of the two features.

Nonetheless, compared with the two cases above in relatively small systems, there are more irregular small bumps in the changing curve, especially in the period when load continuously increases or decreases. Hence with the system scale becoming large, our method may be slightly subjected to complex physical relationship, causing little local disturbances that reduce the effectiveness of our method. However, the overall performance of our method to monitor voltage stability is still of the quality.

*4.4. Operation Speed*

Since the proposed method is intended to be used in real-time, computational time for providing VCP to operators is needed to be discussed. We list the operating time for online calculation in different systems, as shown in Tab. 4. The operation is on the Nvidia GeForce GTX 1080 (8G) GPU. From Tab. 4, the online operating time is very acceptable after proper offline training.

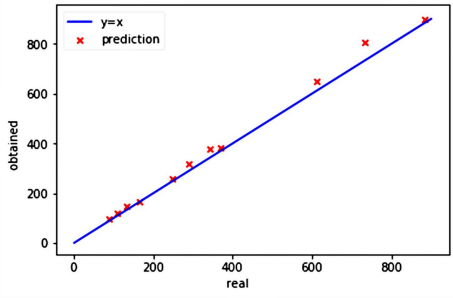
(a) IEEE 14-bus, SLI

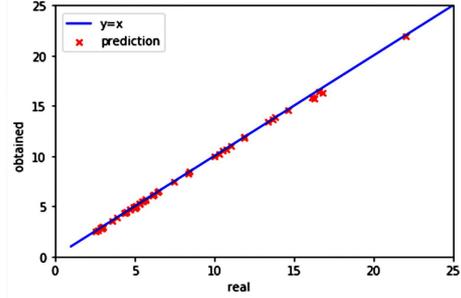
(b) IEEE 14-bus, DLI

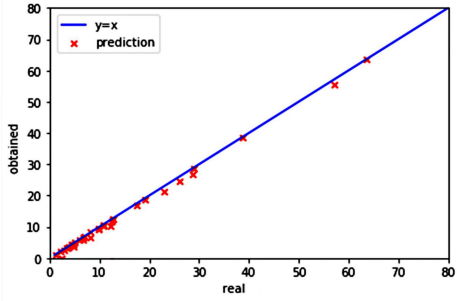
(c) IEEE 57-bus, SLI

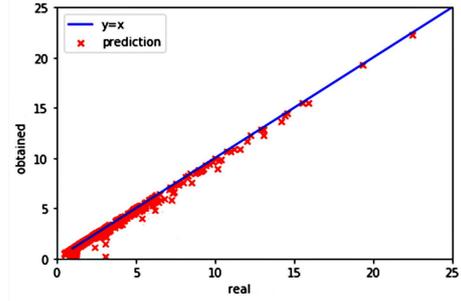
(d) IEEE 57-bus, DLI

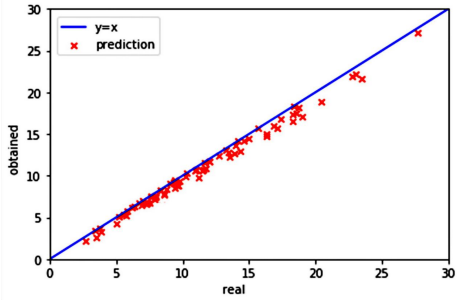
(e) IEEE 118-bus, SLI

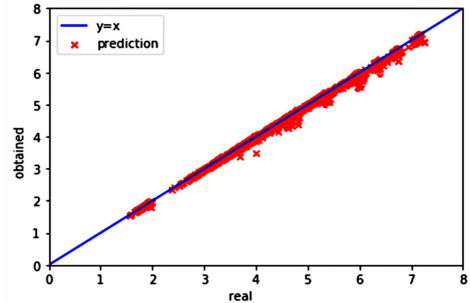
(f) IEEE 118-bus, DLI

Figure 12: The accuracy of VAE. X-aixs and Y-axis denote the real VCP and estimated VCP separately. Blue line is $y = x$. Six figures demonstrate the results of simulations of different power systems and load increment models.

*4.5. Case of VCP Estimation*

A very important point for an indicator to assess or monitor voltage stability is that whether it can clearly reveal VCP. In this case, the accuracy of the proposed index to indicate VCP is tested. In order to compare our approach in different load increment models, we also simulate single load increment (SLI) and double load increment (DLI) cases on all active power load. For simplicity, PMU in IEEE 14-bus, 57-bus and 118-bus system is deployed by the same layout as the above cases.

In this paper, we use mean absolute percent error (MAPE) to measure the prediction error, whose definition is:

$$MAPE = \frac{1}{m} \sum_{i=1}^{m} \frac{|\lambda_{pre,i} - \lambda_{real,i}|}{\lambda_{real,i}} \qquad (15)$$

where $\lambda_{pre,i}$ and $\lambda_{real,i}$ denote the estimated VCP and the real VCP in the *ith* experiment, $m$ is the number of experiments.

To show detailed estimation results, we draw Fig. 12, in which the red crosses plotted at ($\lambda_{real}$, $\lambda_{pre}$) are expected to approach the blue line $y = x$. This six subfigures in Fig. 12 and Tab. 5 demonstrate the effectiveness of our approach.

The accuracy of DLI situation is better than that of SDI situation, illustrating that single variable change is a more extreme situation for power grids in the view of mathematics, as well as more difficult to evaluate voltage stability by the proposed approach.

Table 5: MAPE of VSM estimation results

| MAPE | 14-bus | 57-bus | 118-bus |
|---|---|---|---|
| *SLI* | 0.0341 | 0.0591 | 0.0488 |
| *DLI* | 0.0082 | 0.0338 | 0.0193 |

## 5. Conclusion

This paper proposes a novel online long-term voltage stability assessment and monitoring approach based on VAE. Our method conducts high-dimensional analysis for phasor monitoring voltage data in a probabilistic way. Through analyzing voltage data, two representative features are obtained to evaluate the load level and voltage level respectively, and then to monitor voltage stability. A most important advantage of our approach is that the changing trends are analogous to the real P-V curve. Multiple cases demonstrate the effectiveness and accuracy of our method. In the future, the explicit mathematical derivations of how to search the most representative features in an unsupervised learning way will be our main work.